\documentclass{article}

\usepackage{arxiv}

\usepackage[utf8]{inputenc} 
\usepackage[T1]{fontenc}    
\usepackage{hyperref}       
\usepackage{url}            
\usepackage{booktabs}       
\usepackage{amsfonts}       
\usepackage{amsmath}
\usepackage{nicefrac}       
\usepackage{microtype}      
\usepackage{cleveref}       
\usepackage{lipsum}         
\usepackage{graphicx}
\usepackage{natbib}
\usepackage{doi}
\usepackage{caption}
\usepackage{subcaption}
\usepackage{tikz-cd}

\usepackage{setspace}
\usepackage{algorithm}
\usepackage{algpseudocode}

\newcommand{\stateTuple}[4]{(#1, #2, #3, #4)}

\def\tightlist{}

\title{Dynamic guessing for Hamiltonian Monte Carlo with embedded
numerical root-finding}

\date{}

\usepackage{authblk}

\newbox{\orcid}\sbox{\orcid}{\includegraphics[scale=0.06]{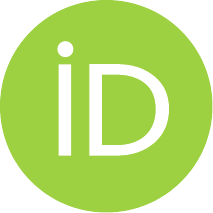}}

\author[1]{
  \href{0000-0002-7109-3270}{\usebox{\orcid}\hspace{1mm}Teddy Groves}
}
\author[1]{
  \href{0000-0003-3367-2105}{\usebox{\orcid}\hspace{1mm}Nicholas Luke
Cowie}
}
\author[1,2]{
  \href{0000-0001-8191-3511}{\usebox{\orcid}\hspace{1mm}Lars Keld
Nielsen}
}

\affil[1]{The Novo Nordisk Center for Biosustainability, DTU, Building
220, Søltofts Plads, Kongens Lyngby, Denmark}
\affil[2]{Australian Institute for Bioengineering and Nanotechnology
(AIBN), The University of Queensland, St Lucia 4067, Australia}

\hypersetup{
	pdftitle={Dynamic guessing for Hamiltonian Monte Carlo with embedded
numerical root-finding},
	pdfauthor={ Teddy Groves,   Nicholas Luke Cowie,   Lars Keld
Nielsen,   },
    colorlinks=true,
    linkcolor=black,
    filecolor=black,
    urlcolor=black,
	citecolor=black
}

\usepackage{color}
\usepackage{fancyvrb}

\DefineVerbatimEnvironment{Highlighting}{Verbatim}{commandchars=\\\{\}}
\newenvironment{Shaded}{}{}

\newcommand{\BuiltInTok}[1]{\textcolor[rgb]{0.00,0.50,0.00}{#1}}

\newcommand{\CommentTok}[1]{\textcolor[rgb]{0.38,0.63,0.69}{\textit{#1}}}

\newcommand{\ControlFlowTok}[1]{\textcolor[rgb]{0.00,0.44,0.13}{\textbf{#1}}}

\newcommand{\DecValTok}[1]{\textcolor[rgb]{0.25,0.63,0.44}{#1}}

\newcommand{\ImportTok}[1]{\textcolor[rgb]{0.00,0.50,0.00}{\textbf{#1}}}

\newcommand{\KeywordTok}[1]{\textcolor[rgb]{0.00,0.44,0.13}{\textbf{#1}}}
\newcommand{\NormalTok}[1]{#1}
\newcommand{\OperatorTok}[1]{\textcolor[rgb]{0.40,0.40,0.40}{#1}}

\usepackage{graphicx}
\makeatletter
\newsavebox\pandoc@box
\newcommand*\pandocbounded[1]{
  \sbox\pandoc@box{#1}%
  \Gscale@div\@tempa{\textheight}{\dimexpr\ht\pandoc@box+\dp\pandoc@box\relax}%
  \Gscale@div\@tempb{\linewidth}{\wd\pandoc@box}%
  \ifdim\@tempb\p@<\@tempa\p@\let\@tempa\@tempb\fi
  \ifdim\@tempa\p@<\p@\scalebox{\@tempa}{\usebox\pandoc@box}%
  \else\usebox{\pandoc@box}%
  \fi%
}
\def\fps@figure{htbp}
\makeatother

\begin{document}
\maketitle

\begin{abstract}
	Modern implementations of Hamiltonian Monte Carlo and related MCMC
algorithms support sampling of probability functions that embed
numerical root-finding algorithms, thereby allowing fitting of
statistical models involving analytically intractable algebraic
constraints. However the application of these models in practice is
limited by the computational cost of computing large numbers of
numerical solutions. We identify a key limitation of previous approaches
to HMC with embedded root-finding, which require the starting guess to
be the same at all points on the same simulated Hamiltonian trajectory.
We demonstrate that this requirement can be relaxed, so that the
starting guess depends on the previous integrator state. To choose a
good guess using this information we propose two heuristics: use the
previous solution and extrapolate the previous solution using implicit
differentiation. Both heuristics yield substantial performance
improvements on a range of representative models compared with static
guessing. We also present \texttt{grapevine}, a JAX-based Python package
providing easy access to an implementation of the No-U-Turn sampler
augmented with dynamic guessing.
\end{abstract}

\section{Introduction}\label{introduction}

If a modeller knows that certain quantities jointly satisfy algebraic
constraints, they may want to make a Bayesian statistical model with an
embedded root-finding problem. To do this, they can express the
unconstrained quantities as model parameters \(\theta\) and then, given
some values of \(\theta\) find the values \(x\) of the constrained
quantities by solving an equation with the form \(f(x, \theta)=0\). If
the function \(f\) expressing the constraints is analytically
intractable, the root-finding problem can often be solved approximately
using numerical methods.

However, HMC sampling of statistical models with algebraic constraints
is often slow due to the computational overhead imposed by numerical
root-finding algorithms. Many applications of Bayesian statistical
inference are currently infeasible due to this issue. For example,
Bayesian statistical models with embedded mechanistically accurate
kinetic models of steady state cellular metabolism are currently limited
to small metabolic pathways of around 20 reactions
\citep{grovesBayesianRegressionFacilitates2024}.

This performance bottleneck arises due to the need to solve the
root-finding problem and find its gradients many times during the course
of HMC sampling. To generate a single posterior sample, the sampler must
evaluate the log probability density and its parameter gradients,
solving the embedded root-finding problem along the way, at each step
along a simulated Hamiltonian trajectory. To illustrate, a reasonably
complicated model may on average require 500 steps per trajectory; in
this case, obtaining 2000 posterior samples requires solving the
embedded root-finding problem one million times.

In this paper we identify a key limitation of previous implementations
of HMC with embedded root-finding: these implementations require that
the embedded root-finding algorithm use the same starting guess for all
problems that lie on the same simulated Hamiltonian trajectory. We
propose relaxing this requirement, allowing the integrator to update the
starting guess dynamically based on the previous state of the sampler.
We propose two heuristics for choosing a new starting guess. The first
heuristic simply uses the previous solution as the next starting guess.
The second heuristic perturbs the previous guess based on the local
change in parameter space using implicit differentiation. We test both
of these heuristics on a range of models, showing that dynamic guessing
improves performance compared with the state of the art, and that the
best heuristic depends on the specific problem. We also present a Python
package \texttt{grapevine} containing our implementation of HMC with
dynamic guessing, benchmarks and convenience functions that allow
application of dynamic guessing to arbitrary statistical models.

\section{Related work}\label{related-work}

Bayesian statistical models with embedded analytically intractable
root-finding problems are reported in a wide range of scientific
contexts, including ignition chemistry
\citep{najmUncertaintyQuantificationChemical2009}, cell biology
\citep{grovesBayesianRegressionFacilitates2024, linden-santangeliIncreasingCertaintySystems2024}
and optimal control
\citep{leeftinkProbabilisticPontryaginsMaximum2025a}.

Hamiltonian Monte Carlo
\citep{nealMCMCUsingHamiltonian2011, betancourtConceptualIntroductionHamiltonian2018}
and related algorithms such as the No-U-Turn sampler
\citep{hoffmanNoUTurnSamplerAdaptively2014} (unless specified otherwise,
we include such variants under ``HMC'') support sampling for this kind
of model: see
\citep{timonenImportanceSamplingApproach2022, margossianReviewAutomaticDifferentiation2019}.
HMC is often preferable to alternative inference algorithms because of
its good performance \citep{mangoubiDoesHamiltonianMonte2018},
asymptotic exactness and the existence of highly-used and
well-maintained implementations, e.g.
\citep{carpenterStanProbabilisticProgramming2017, abril-plaPyMCModernComprehensive2023}.

Hamiltonian Monte Carlo couples the target probability distribution with
a dynamical system representing a particle lying on a surface with one
dimension per parameter of the target distribution. To generate a
sample, a symplectic integrator simulates the particle's trajectory
along the surface after a perturbation. Ideally, the trajectory takes
the particle far from its starting point, leading to efficient sampling.
The symplectic integrator proceeds by linearising the trajectory in
small segments, evaluating the probability density and its parameter
gradients on log scale at the end of each segment. For a target
distribution with embedded numerical root-finding, the roots and their
local parameter gradients must also be found at every one of these
steps.

Many popular numerical root-finding algorithms are iterative, generating
a series of numbers that start with an initial guess and converge
asymptotically towards the true solution. These algorithms tend to
perform better, the closer the initial guess is to the true solution:
see \citep{casellaChoiceInitialGuesses2021} for discussion of this
topic. Thus a natural way to speed up HMC with embedded root-finding is
to find the best possible guess for each problem; indeed, the main
recommendation of the Stan user guide
\citep{standevelopmentteamStanModelingLanguage2025} is to use a global
guess that is reasonable, given the likely values of the parameters.
However, since HMC trajectories aim to traverse a large distance in
parameter space, and embedded root-finding problems will typically have
different solutions depending on the parameters, the solution is likely
to vary for different points on the trajectory. As a result, the choice
of initial guess must take into account a range of solutions.

\section{Methods}\label{methods}

Instead of using the same starting guess for every root-finding problem
on the same simulated Hamiltonian trajectory, we propose choosing the
guess dynamically, based on the previous integrator state. We call this
approach the ``grapevine method''.

Specifically, we propose augmenting the Velocity Verlet integrator
\citep{swopeComputerSimulationMethod1982, blanesNumericalIntegratorsHybrid2014}
with a dynamic state variable containing information that will be used
to calculate an initial guess for the root-finding algorithm when the
integrator updates its position. This variable is initialised at a
default value, which is used to solve the numerical problem at the first
step of any trajectory, and then modified at each position update. In
this way all steps except the first have access to non-default guessing
information.

The algorithm is described below in pseudocode.

\begin{algorithm}[H]
\caption{Generate a new guess (Heuristic)}
\label{alg:heuristic}
\begin{algorithmic}[1]\onehalfspacing
\Require Information from previous step, $\operatorname{info}$.
\Ensure  A guess for the root-finding problem, $x_{guess}$.
\State \Return $x_{guess}$ based on $\operatorname{info}$. \Comment{e.g., return solution from previous step}
\end{algorithmic}
\end{algorithm}
\begin{algorithm}[H]
\caption{Update log density and information (LogDensityAndInfo)}
\label{alg:logdensity}
\begin{algorithmic}[1]\onehalfspacing
\Require Parameters $\theta$; information from previous step, $\operatorname{info}$.
\Ensure  Log probability density $lp$; updated information $\operatorname{info}_{\text{next}}$.
\State $x_{guess} \gets \text{Heuristic}(\operatorname{info})$ \Comment{Get initial guess for the algebraic problem}
\State $x_{curr} \gets \text{Solve}(f(x, \theta) = 0, \text{starting from } x_{guess})$ \Comment{Solve embedded algebraic equation}
\State $lp \gets \text{LogProb}(\theta, x_{curr})$ \Comment{Calculate log probability density}
\State $\operatorname{info}_{\text{next}} \gets (x_{curr}, \theta, ...)$ \Comment{Store information for the next step}
\State \Return $lp, \operatorname{info}_{\text{next}}$
\end{algorithmic}
\end{algorithm}
\begin{algorithm}[H]
\caption{Evaluate potential energy and gradient (PotentialAndGradient)}
\label{alg:potential}
\begin{algorithmic}[1]\onehalfspacing
\Require Parameters $\theta$; information from previous step, $\operatorname{info}$.
\Ensure  Potential energy $U$, updated information $\operatorname{info}_{\text{next}}$, and gradient $\nabla_{\theta} U$.
\State $lp, \operatorname{info}_{\text{next}} \gets \text{LogDensityAndInfo}(\theta, \operatorname{info})$ \Comment{Compute log density and get next info}
\State $U \gets -lp$ \Comment{Potential energy is the negative log density}
\State $\nabla_{\theta} U \gets -\nabla_{\theta} lp$ \Comment{Compute the gradient of the potential energy}
\State \Return $U, \operatorname{info}_{\text{next}}, \nabla_{\theta} U$
\end{algorithmic}
\begin{algorithm}[H]
\caption{Trajectory Initialization}
\label{alg:init}
\begin{algorithmic}[1]\onehalfspacing
\Require Initial parameters $\theta_0$; initial momentum $p_0$; default guess information $x_{\text{default\_guess}}$.
\Ensure  Initial state $(\theta_0, p_0, \nabla_{\theta} U(\theta_0), \operatorname{info}_0)$
\State $U(\theta_0), \operatorname{info}_0, \nabla_{\theta} U(\theta_0) \gets \text{PotentialAndGradient}(\theta_0, x_{\text{default\_guess}})$ \Comment{Get initial potential, gradient and info}
\State \textbf{return} $\stateTuple{\theta_0}{p_0}{\nabla_{\theta} U(\theta_0)}{\operatorname{info}_0}$ \Comment{The full state is returned to be used by the first leapfrog step}
\end{algorithmic}
\end{algorithm}
\end{algorithm}
\begin{algorithm}[H]
\caption{Leapfrog Integration Step}
\label{alg:leapfrog}
\begin{algorithmic}[1]\onehalfspacing
\Require Current state $(\theta, p, \nabla_{\theta} U(\theta), \operatorname{info}_{\text{prev}})$; step size $\epsilon$.
\Ensure  Updated state $(\theta_{\text{next}}, p_{\text{next}}, \nabla_{\theta} U(\theta_{\text{next}}), \operatorname{info}_{\text{next}})$.
\State $p_{\text{mo}} \gets p - \frac{\epsilon}{2} \nabla_{\theta} U(\theta)$ \Comment{Update momentum (first half-step)}
\State $\theta_{\text{next}} \gets \theta + \epsilon p_{\text{mo}}$ \Comment{Update parameters}
\State $U(\theta_{\text{next}}), \operatorname{info}_{\text{next}}, \nabla_{\theta} U(\theta_{\text{next}}) \gets \text{PotentialAndGradient}(\theta_{\text{next}}, \operatorname{info}_{\text{prev}})$ \Comment{Update potential, info and gradient}
\State $p_{\text{next}} \gets p_{\text{mo}} - \frac{\epsilon}{2} \nabla_{\theta} U(\theta_{\text{next}})$ \Comment{Update momentum (second half-step)}
\end{algorithmic}
\end{algorithm}

\subsection{Heuristics}\label{heuristics}

We propose two heuristics for generating an initial guess:
\emph{guess-previous} and \emph{guess-implicit}.

The information for the heuristic \emph{guess-previous} is the solution
\(x_{prev}\) of the previous root-finding problem. The heuristic is
simply to use the previous solution as the next guess:

\[
\text{\emph{guess-previous}}(x_{prev}) = x_{prev}
\]

The information for the heuristic \emph{guess-implicit} is the solution
\(x_{prev}\) of the previous root-finding problem, and also the
parameters \(\theta_{prev}\) of the previous step. The heuristic is to
use implicit differentiation to find \(\frac{dx}{d\theta}\), the local
derivative of \(x\) with respect to \(\theta\), then obtain a guess
according to this formula:

\[
\text{\emph{guess-implicit}}(x_{prev}, \theta_{prev}, \theta_{next}) = x_{prev} + \frac{dx}{d\theta}\nabla_{\theta}
\]

where \(\nabla_{\theta}=(\theta_{next}-\theta_{prev})\). To obtain
\(\frac{dx}{d\theta}\), we use the following consequence of the implicit
function theorem\citep{oliveiraImplicitInverseFunction2014}:

\[
\frac{\delta x}{\delta\theta} = -(\operatorname{jac}_{x}f(x_{prev}, \theta_{prev}))^{-1}\operatorname{jac}_{\theta}f(x_{prev}, \theta_{prev})
\]

In this expression the term
\(\operatorname{jac}_{x}f(x_{prev}, \theta_{prev})\), abbreviated below
to \(J_{x}\), indicates the jacobian with respect to \(x\) of
\(f(x_{prev}, \theta_{prev})\). Similarly
\(\operatorname{jac}_{\theta}f(x_{prev}, \theta_{prev}) = J_{\theta}\)
is the jacobian with respect to \(\theta\) of
\(f(x_{prev}, \theta_{prev})\).

Substituting terms we then have

\[
\text{\emph{guess-implicit}}(x_{prev}, \theta_{prev}, \theta_{next}) = x_{prev} - J_{x}^{-1}J_{\theta} \nabla_{\theta}
\]

The \emph{guess-implicit} heuristic can be implemented using the
following Python function:

\begin{Shaded}
\begin{Highlighting}[]
\ImportTok{import}\NormalTok{ jax }

\KeywordTok{def}\NormalTok{ guess\_implicit(guess\_info, params, f):}
    \CommentTok{"Guess the next solution using the implicit function theorem."}
\NormalTok{    old\_x, old\_p, }\OperatorTok{*}\NormalTok{\_ }\OperatorTok{=}\NormalTok{ guess\_info}
\NormalTok{    delta\_p }\OperatorTok{=}\NormalTok{ jax.tree.}\BuiltInTok{map}\NormalTok{(}\KeywordTok{lambda}\NormalTok{ o, n: n }\OperatorTok{{-}}\NormalTok{ o, old\_p, params)}
\NormalTok{    \_, jvpp }\OperatorTok{=}\NormalTok{ jax.jvp(}\KeywordTok{lambda}\NormalTok{ p: f(old\_x, p), (old\_p,), (delta\_p,))}
\NormalTok{    jacx }\OperatorTok{=}\NormalTok{ jax.jacfwd(f, argnums}\OperatorTok{=}\DecValTok{0}\NormalTok{)(old\_x, old\_p)}
\NormalTok{    u }\OperatorTok{=} \OperatorTok{{-}}\NormalTok{(jnp.linalg.inv(jacx))}
    \ControlFlowTok{return}\NormalTok{ old\_x }\OperatorTok{+}\NormalTok{ u }\OperatorTok{@}\NormalTok{ jvpp}
\end{Highlighting}
\end{Shaded}

Note that this function avoids materialising the parameter jacobian
\(J_{\theta}\), instead finding the jacobian vector product
\(J_{\theta}\nabla_{\theta}\) using the function \texttt{jax.jvp}. It is
possible to avoid materialising the matrix \(J_{x}\) using a similar
strategy, as demonstrated by the function \texttt{guess\_implicit\_cg}
below.

\begin{Shaded}
\begin{Highlighting}[]
\ImportTok{import}\NormalTok{ jax }

\KeywordTok{def}\NormalTok{ guess\_implicit\_cg(guess\_info, params, f):}
    \CommentTok{"Guess the next solution using the implicit function theorem."}
\NormalTok{    old\_x, old\_p, }\OperatorTok{*}\NormalTok{\_ }\OperatorTok{=}\NormalTok{ guess\_info}
\NormalTok{    delta\_p }\OperatorTok{=}\NormalTok{ jax.tree.}\BuiltInTok{map}\NormalTok{(}\KeywordTok{lambda}\NormalTok{ o, n: n }\OperatorTok{{-}}\NormalTok{ o, old\_p, params)}
\NormalTok{    \_, jvpp }\OperatorTok{=}\NormalTok{ jax.jvp(}\KeywordTok{lambda}\NormalTok{ p: f(old\_x, p), (old\_p,), (delta\_p,))}

    \KeywordTok{def}\NormalTok{ matvec(v):}
        \CommentTok{"Compute Jx @ v"}
        \ControlFlowTok{return}\NormalTok{ jax.jvp(}\KeywordTok{lambda}\NormalTok{ x: f(x, old\_p), (old\_x,), (v,))[}\DecValTok{1}\NormalTok{]}

\NormalTok{    dx }\OperatorTok{=} \OperatorTok{{-}}\NormalTok{jax.scipy.sparse.linalg.cg(matvec, jvpp)[}\DecValTok{0}\NormalTok{]}
    \ControlFlowTok{return}\NormalTok{ old\_x }\OperatorTok{+}\NormalTok{ dx}
\end{Highlighting}
\end{Shaded}

Which implementation of \emph{guess-implicit} is preferable depends on
the relative cost and reliability of directly calculating the matrix
inverse \(J_{x}^{-1}\) as in the function \texttt{guess\_implicit},
compared with numerically solving \(J_x J_p \nabla_p = 0\) as in
\texttt{guess\_implicit\_cg}. In general, this depends on the
performance of the characteristics of the numerical solver relative to
direct matrix inversion as implemented by the function
\texttt{jax.numpy.linalg.inv}. For example, if \(J_{x}\) is sparse but
positive semi-definite, \texttt{guess\_implicit\_cg} will likely perform
better as the conjugate gradient method can exploit sparsity
\citep{ben-talCONVEXANALYSISNONLINEAR}.

\subsection{Implementation}\label{implementation}

We augmented the velocity Verlet integrator
\citep{blanesNumericalIntegratorsHybrid2014} provided by Blackjax
\citep{cabezas2024blackjax} with a dynamic guessing variable, and used
this new integrator to create a No-U-Turn sampler with dynamic guessing,
which we call ``grapeNUTS''. For convenience we provide a Python package
\texttt{grapevine} containing our implementation, including a utility
function \texttt{run\_grapenuts} with which users can easily test the
GrapeNUTS sampler. See the code repository file \texttt{README.md} for
installation and usage instructions.

Our implementation builds on the popular JAX \citep{jax2018github}
scientific computing ecosystem, allowing users to straightforwardly
define statistical models and adapt existing models to work with
grapeNUTS. Similarly to Blackjax and Bayeux
\citep{bayeuxdevelopersBayeuxStateArt2025}, grapevine requires a model
in the form of a function that returns a scalar log probability density
given a JAX PyTree of parameters; additionally, in grapevine such a
function must also accept and return a PyTree containing information to
be used for guessing the answers to embedded root-finding problems.
Users can specify root-finding problems using arbitrary JAX-compatible
libraries, for example optimistix \citep{optimistix2024} or diffrax
\citep{kidger2021on}.

Thanks to the modular design of JAX, Blackjax and optimistix, users can
straightforwardly extend our implementation to create grapevine versions
of other symplectic integrators and MCMC algorithms.

\subsection{Code and data
availability}\label{code-and-data-availability}

Our implementation of the grapevine method, and the code used to perform
the experiment results reported in this paper, are available at
\url{https://github.com/dtu-qmcm/grapevine}.

\section{Results}\label{results}

\subsection{Illustration}\label{illustration}

\begin{figure}[htbp]
\centering
\includegraphics[width=0.8\textwidth, height=!]{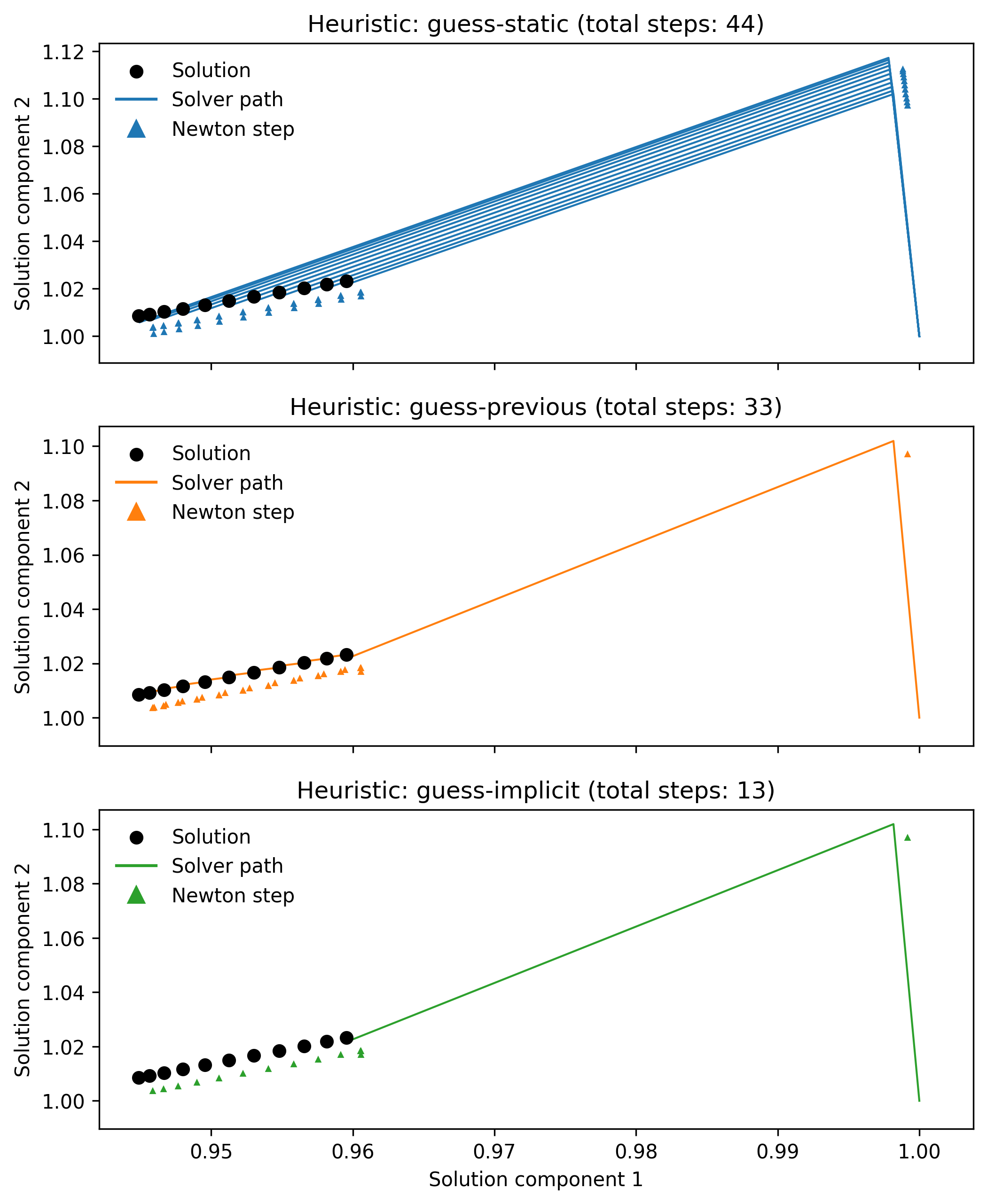}
\caption{An example illustrating the benefit of dynamic guessing by comparing behaviour of dynamic and static heuristics along a single Hamiltonian trajectory. Black dots show the solution to the root-finding problem of finding the minimum of a parametrised 2-dimensional Rosenbrock function, at each of 11 steps along a simulated Hamiltonian trajectory through parameter space. Note that the x and y axis show solution components not parameter components. Coloured crosses show intermediate solutions from a Newton solver. Lines show the path from step to step. To find the first, rightmost problem on the trajectory, all heuristics use the default guess at coordinate (1, 1). For subsequent problems, the \textit{guess-static} heuristic (blue lines) continues to use the default guess, the \textit{guess-previous} heuristic (orange lines) guesses the previous solution and the \textit{guess-implicit} heuristic (green lines) finds a guess using implicit differentiation of the previous solution and parameters. The dynamic heuristics produce better guesses in this case, resulting in shorter paths and fewer Newton steps required to solve all root-finding problems on the trajectory.  See the code repository files benchmarks/trajectory.py and benchmarks/analyse\_results.py for code that generated this plot.}
\label{fig-trajectory}
\end{figure}

Figure \ref{fig-trajectory} illustrates how the grapevine method works
by plotting how our two proposed heuristics behave compared with static
guessing when solving embedded root-finding problems along a single
Hamiltonian trajectory.

\subsection{Experiments}\label{experiments}

\begin{figure}[htbp]
\centering
\includegraphics[width=0.8\textwidth]{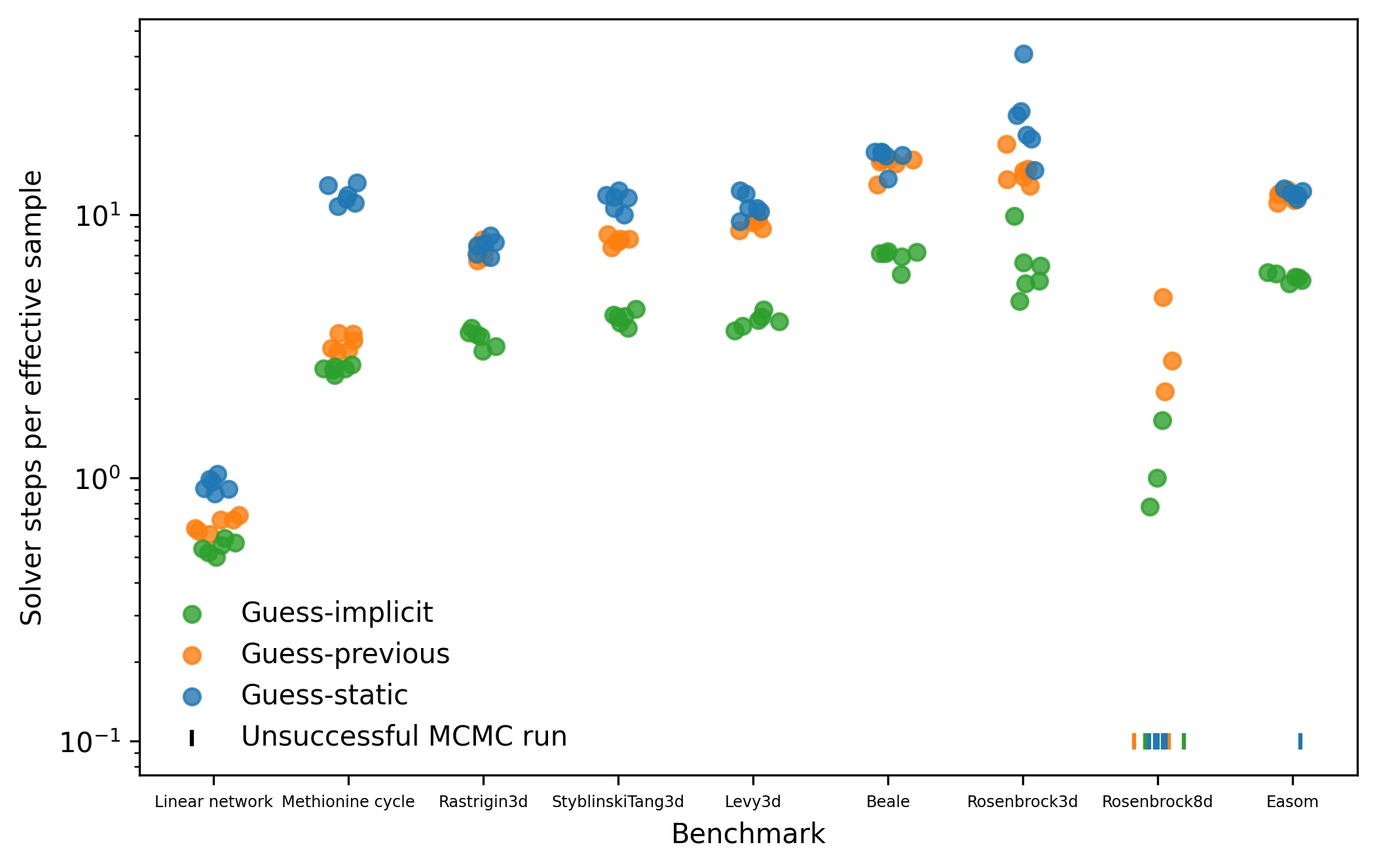}
\caption{Performance comparison for three dynamic guessing heuristics over nine statistical models with embedded root-finding problems. For each point, a true parameter set was randomly selected, then a simulated dataset was generated consistently with the target model and used to generate posterior samples. Sampler performance is quantified by the number of effective samples generated divided by the total number of solver steps, as plotted on logarithmic scale on the y axis. Lower values indicate better performance. Vertical lines denote MCMC runs that were unsuccessful because of a numerical solver failure. Note that  See the code repository file \texttt{benchmarks/analyse\_results.py} for code used to generate this plot.}
\label{fig-benchmarks}
\end{figure}

Figure \ref{fig-benchmarks} shows the results of our experiments. These
were performed on a MacBook Pro 2024 with Apple M4 Pro processor and
48GB RAM, running macOS 15.3.1.

We compared the performance of dynamic vs static guessing by fitting a
range of models with embedded root-finding problems using our
implementation of GrapeNUTS. The code that carried out these experiments
is in the code repository directory \texttt{benchmarks}: instructions
for reproducing the experiments are in the file \texttt{README.md}.

For each model, we randomly generated 6 parameter sets, each of which we
used to randomly simulate one fake observation set. We then sampled from
the resulting posterior distribution using grapeNUTS and each of four
guessing heuristics: \emph{guess-previous}, \emph{guess-implicit} with
materialised jacobian, \emph{guess-implicit} with non-materialised
jacobian and a dummy guessing heuristic implementing static guessing.
The static guessing heuristic is equivalent to NUTS with no
augmentation, therefore providing a convenient baseline. The sampler
configurations were the same for all heuristics. As a solver we used the
Newton solver provided by optimistix, with the tolerances set based on
each problem.

We quantified performance by calculating using the total wall time and
the total number of Newton steps. Sampling was diagnosed by inspecting
the effective sample size using arviz
\citep{kumarArviZUnifiedLibrary2019}, verifying that this quantity was
not small compared with the total number of MCMC samples, and by
verifying that there were no post-warmup divergent transitions. We also
recorded MCMC runs when the solver failed to converge at least once - in
these cases we counted the whole run as a failure.

\subsubsection{Models}\label{models}

\subparagraph{Optimisation test
functions}\label{optimisation-test-functions}

We compared our four heuristics on a series of variations of the
following model:

\begin{align*}
\theta &\sim Normal(0, \sigma_{\theta}) \\
\hat{y} &= root_y(f(y + \theta)) \\
y &\sim Normal(\hat{y}, \sigma_{y})
\end{align*}

In this equation \(f\) is the gradient of a textbook optimisation test
function, \(sol\) is the textbook solution and \(\theta\) is a vector
with the same size as the input to \(f\). We tested the following
functions from the virtual library of simulation experiments
\citep{simulationlib}:

\begin{itemize}
\tightlist
\item
  Easom function (2 dimensions)
\item
  3-dimension Levy function
\item
  Beale function (2 dimensions)
\item
  3-dimension Rastrigin function
\item
  3-dimension and 8-dimension Rosenbrock functions
\item
  Styblinski-Tang function
\end{itemize}

We chose these functions because they have a range of different
difficult features for numerical solvers, vary in dimensions, have
global minima so that the associated root-finding problems are
well-posed and are straightforward to implement.

\subsubsection{Steady-state reaction
networks}\label{steady-state-reaction-networks}

To illustrate our algorithm's practical relevance we constructed two
statistical models where evaluating the likelihood \(p(y\mid\theta)\)
requires solving a steady state problem, i.e.~finding a vector \(x\)
such that \(\frac{dx}{dt} = S\cdot v(x, \theta) = \bar{0}\) for known
real-valued matrix \(S\) and function \(v\). In the context of chemical
reaction networks, \(S_{ij}\in\mathbb{R}\) can be interpreted as
representing the amount of compound \(i\) consumed or produced by
reaction \(j\), \(x\) as the abundance of each compound and
\(v(x, \theta)\) as the rate of each reaction. The condition
\(\frac{dx}{dt} = \bar{0}\)then represents the assumption that the
compounds' abundances are constant. This kind of model is common in many
fields, especially biochemistry: see for example
\citep{matosGRASPComputationalPlatform2022, fiedlerTailoredParameterOptimization2016, grovesBayesianRegressionFacilitates2024}.

We tested two similar models with this broad structure, one embedding a
small biologically-inspired steady state problem and one a relatively
large and well-studied realistic steady state problem.

The smaller modelled network is a toy model of a linear pathway with
three reversible reactions with rates \(v_1\), \(v_2\) and \(v_3\).
These reactions affect the internal concentrations \(A^{int}\) and
\(B^{int}\) according to the following graph:

\begin{center}
  \begin{tikzcd}[column sep=small]
    &&&&&&&&&&&& {} \\
    & {\ } &&&&&& {\ } \\
    {A^{ext}} &&& {A^{int}} && {B^{int}} &&& {B^{ext}} \\
    & {\ } &&&&&& {\ }
    \arrow[dashed, no head, from=2-2, to=4-2]
    \arrow[dashed, no head, from=2-8, to=2-2]
    \arrow["{v_1}", <->, from=3-1, to=3-4]
    \arrow["{v_2}", <->, from=3-4, to=3-6]
    \arrow["{v_3}", <->, from=3-6, to=3-9]
    \arrow[dashed, no head, from=4-2, to=4-8]
    \arrow[dashed, no head, from=4-8, to=2-8]
  \end{tikzcd}
\end{center}

The rates \(v_1\), \(v_2\) and \(v_3\) are calculated as follows, given
internal concentrations \(x^{int} = x^{int}_{A}, x^{int}_{B}\) and
parameters \(\theta
= k^{m}_{A}, k^{m}_{B}, v^{max}, k^{eq}_1, k^{eq}_2,k^{eq}_3, k^{f}_1, k^{f}_3, x^{ext}_A, x^{ext}_{B}\):

\[
\begin{aligned}
v_1(x^{int}, \theta) &=  k^{f}_1 (x^{ext}_{A} - x^{int}_{A} / k^{eq}_1) \\
v_2(x^{int}, \theta) &= \frac{\frac{v^{max}}{k^{m}_A} (x^{int}_{A} - x^{int}_{B} / k^{eq}_2)}{1 + x^{int}_{A}/k^{m}_{A} + x^{int}_{B}/k^{m}_{B} }  \\
v_3(x^{int}, \theta) &=  k^{f}_3 (x^{ext}_{B} - x^{int}_{B} / k^{eq}_3)
\end{aligned}
\]

According to these equations, rates \(v_1\) and \(v3\) described by
mass-action rate laws: transport reactions are often modelled in this
way. Rate \(v_2\) is described by the Michaelis-Menten equation that is
a popular choice for modelling the rates of enzyme-catalysed reactions.

The larger network models the mammalian methionine cycle, using
equations taken from \citep{grovesBayesianRegressionFacilitates2024},
including highly non-linear regulatory interactions. We selected this
model because it describes a real biological system and has a convenient
scale, being large and complex enough to test the grapevine method's
scalability, but small enough for benchmarking purposes.

For the small linear network, we solved the embedded steady state
problem using the optimistix Newton solver. For the larger model of the
methionine cycle we simulated the evolution of internal concentrations
as an initial value problem until a steady state event occurred, using
the steady state event handler and Kvaerno5 ODE solver provided by
diffrax. In this case a guess is still needed in order to provide an
initial value. Solving a steady state problem in this way is often more
robust than directly solving the system of algebraic equations; see
\citep{fiedlerTailoredParameterOptimization2016} and
\citep[Introduction]{lakrisenkoEfficientComputationAdjoint2023} for
further discussion.

Code used for these two experiments is in the code repository files
\texttt{benchmarks/methionine.py} and \texttt{benchmarks/linear.py}.

\section{Discussion and conclusions}\label{discussion-and-conclusions}

Dynamic guessing tended to improve MCMC performance compared with static
guessing for all the statistical models that we tested. The heuristic
\emph{guess-previous} showed similar or better performance compared with
\emph{guess-static}, whereas \emph{guess-implicit} performed
substantially better than \emph{guess-static} on every benchmark.

It is also notable that the dynamic algorithms failed less frequently
than \emph{guess-static} on the difficult \texttt{Rosenbrock8d} and
\texttt{Easom} benchmarks. This is likely because the dynamic algorithms
are less likely to fail when traversing low-probability trajectories at
the start of the adaptation phase. These trajectories are especially
unfavourable for static guessing because they have root-finding problem
solutions that are far away from any reasonable global guess.

Based on these results, we expect that replacing a static guessing
algorithm with the grapevine method will typically improve sampling
performance for similar MCMC tasks, making it possible to fit previously
infeasible statistical models.

While dynamic guessing generally outperformed static guessing, the
relative performance of \emph{guess-previous} compared with
\emph{guess-implicit} varied between benchmarks. We expect that this
variation was caused by differences between benchmark problems in how
smoothly the solution of the embedded problem changes with changes in
parameters. The smoother this relationship, the more likely that the
embedded problems at adjacent points in an HMC trajectory will have
similar solutions, leading to better relative performance of the
\emph{guess-previous} heuristic.

An opportunity for further performance improvement would be to use a
different method to solve the first root-finding problem in a trajectory
than for later problems. Plausibly, a slow but robust solver could be
preferable for the first problem, which uses a default guess, whereas a
faster but more fragile solver might be preferable for later problems
where a potentially better guess is available.

While our implementation of the grapevine method is performant and
flexible, it must be used with care, as it requires a log posterior
function where the guess variable is only used by the numerical solver,
and does not otherwise affect the output density. In our implementation
there is no automatic safeguard preventing the user from writing a log
probability density function where the output density depends on the
guess, even though doing so risks producing invalid MCMC inference. In
future, it may be beneficial to implement a stricter grapevine interface
that makes inappropriate use of the guess variable impossible.

\section{Acknowledgements}\label{acknowledgements}

This work was funded by the Novo Nordisk Foundation (grant numbers
NNF20CC0035580 NNF14OC0009473).

\bibliographystyle{unsrtnat}
\bibliography{bibliography.bib}
\end{document}